\PassOptionsToPackage{usenames,dvipsnames}{xcolour}
\documentclass[twocolumn,twocolappendix,nofootinbib,iop]{openjournal}

\pdfoutput=1 
\usepackage{amsmath,amssymb,amstext}
\usepackage[T1]{fontenc}
\usepackage{apjfonts}
\usepackage{ae,aecompl}
\usepackage[utf8]{inputenc}
\usepackage[colorlinks,allcolors=blue]{hyperref}
\usepackage[figure,figure*]{hypcap}
\usepackage{natbib}
\usepackage{url}
\usepackage{mdwlist}
\usepackage{multirow}
\usepackage{lineno}
\usepackage{fontawesome}
\usepackage{array}
\usepackage{xcolor}

\usepackage{tabularx}
\newcolumntype{L}{>{\raggedright\arraybackslash}X}

\newcolumntype{L}{>{\raggedright\arraybackslash}X}

\usepackage{soul}

\usepackage{xcolor}
\definecolor{maroon}{cmyk}{0, 0.87, 0.68, 0.32}
\definecolor{halfgray}{gray}{0.55}
\definecolor{ipython_frame}{RGB}{207, 207, 207}
\definecolor{ipython_bg}{RGB}{247, 247, 247}
\definecolor{ipython_red}{RGB}{186, 33, 33}
\definecolor{ipython_green}{RGB}{0, 128, 0}
\definecolor{ipython_cyan}{RGB}{64, 128, 128}
\definecolor{ipython_purple}{RGB}{170, 34, 255}

\usepackage{listings}

\lstdefinelanguage{iPython}{
    morekeywords={access,and,break,class,continue,def,del,elif,else,except,exec,finally,for,from,global,if,import,in,is,lambda,not,or,pass,print,raise,return,try,while},%
    morekeywords=[2]{abs,all,any,basestring,bin,bool,bytearray,callable,chr,classmethod,cmp,compile,complex,delattr,dict,dir,divmod,enumerate,eval,execfile,file,filter,float,format,frozenset,getattr,globals,hasattr,hash,help,hex,id,input,int,isinstance,issubclass,iter,len,list,locals,long,map,max,memoryview,min,next,object,oct,open,ord,pow,property,range,raw_input,reduce,reload,repr,reversed,round,set,setattr,slice,sorted,staticmethod,str,sum,super,tuple,type,unichr,unicode,vars,xrange,zip,apply,buffer,coerce,intern},%
    sensitive=true,%
    morecomment=[l]\#,%
    morestring=[b]',%
    morestring=[b]",%
    morestring=[s]{'''}{'''},
    morestring=[s]{"""}{"""},
    morestring=[s]{r'}{'},
    morestring=[s]{r"}{"},%
    morestring=[s]{r'''}{'''},%
    morestring=[s]{r"""}{"""},%
    morestring=[s]{u'}{'},
    morestring=[s]{u"}{"},%
    morestring=[s]{u'''}{'''},%
    morestring=[s]{u"""}{"""},%
    literate=
    {á}{{\'a}}1 {é}{{\'e}}1 {í}{{\'i}}1 {ó}{{\'o}}1 {ú}{{\'u}}1
    {Á}{{\'A}}1 {É}{{\'E}}1 {Í}{{\'I}}1 {Ó}{{\'O}}1 {Ú}{{\'U}}1
    {à}{{\`a}}1 {è}{{\`e}}1 {ì}{{\`i}}1 {ò}{{\`o}}1 {ù}{{\`u}}1
    {À}{{\`A}}1 {È}{{\'E}}1 {Ì}{{\`I}}1 {Ò}{{\`O}}1 {Ù}{{\`U}}1
    {ä}{{\"a}}1 {ë}{{\"e}}1 {ï}{{\"i}}1 {ö}{{\"o}}1 {ü}{{\"u}}1
    {Ä}{{\"A}}1 {Ë}{{\"E}}1 {Ï}{{\"I}}1 {Ö}{{\"O}}1 {Ü}{{\"U}}1
    {â}{{\^a}}1 {ê}{{\^e}}1 {î}{{\^i}}1 {ô}{{\^o}}1 {û}{{\^u}}1
    {Â}{{\^A}}1 {Ê}{{\^E}}1 {Î}{{\^I}}1 {Ô}{{\^O}}1 {Û}{{\^U}}1
    {œ}{{\oe}}1 {Œ}{{\OE}}1 {æ}{{\ae}}1 {Æ}{{\AE}}1 {ß}{{\ss}}1
    {ç}{{\c c}}1 {Ç}{{\c C}}1 {ø}{{\o}}1 {å}{{\r a}}1 {Å}{{\r A}}1
    {€}{{\EUR}}1 {£}{{\pounds}}1
    {^}{{{\color{ipython_purple}\^{}}}}1
    {=}{{{\color{ipython_purple}=}}}1
    {+}{{{\color{ipython_purple}+}}}1
    {-}{{{\color{ipython_purple}-}}}1
    {*}{{{\color{ipython_purple}$^\ast$}}}1
    {/}{{{\color{ipython_purple}/}}}1
    {+=}{{{+=}}}1
    {-=}{{{-=}}}1
    {*=}{{{$^\ast$=}}}1
    {/=}{{{/=}}}1,
    literate=
    *{-}{{{\color{ipython_purple}-}}}1
     {?}{{{\color{ipython_purple}?}}}1,
    identifierstyle=\color{black}\ttfamily,
    commentstyle=\color{ipython_cyan}\ttfamily,
    stringstyle=\color{ipython_red}\ttfamily,
    keepspaces=true,
    showspaces=false,
    showstringspaces=false,
    rulecolor=\color{ipython_frame},
    frameround={t}{t}{t}{t},
    numbers=none,
    numberstyle=\tiny\color{halfgray},
    backgroundcolor=\color{ipython_bg},
    basicstyle=\ttfamily\footnotesize,
    columns=fullflexible,
    keywordstyle=\color{ipython_green}\ttfamily,
}

\newcommand{\nblink}[1]{\href{https://github.com/DifferentiableUniverseInitiative/jax-cosmo-paper/blob/master/notebooks/#1.ipynb}{\faFileCodeO}}

\begin{document}
\journalinfo{The Open Journal of Astrophysics}

\title{Quantifying the Fermi paradox via passive SETI: a general framework}


\author{
M. Civiletti$^{1,\ast}$
}
\thanks{$^\ast$matthew.civiletti@qc.cuny.edu}

\affiliation{
$^1$Queens College, City University of New York
}


\begin{abstract}
In this paper we consider the extent to which a lack of observations from SETI may be used to quantify the Fermi paradox. We compute the probability of at least one detection of an extraterrestrial electromagnetic (EM) signal of Galactic origin, as a function of the number $N$ of communicative civilizations. We show how this is derivable from the probability of detecting a single signal; the latter is $\approx 0.6 \delta/R$, where $\delta$ is the distance between the initial and final EM signals and $R$ is the radius of the Milky Way, for $\delta/R \ll 1$. We show how to combine this analysis with the Drake equation $N = \mathcal{N} \delta /c$, where $c$ is the speed of light; this implies, applying a simplified toy model as an example, that the probability of detecting at least one signal is $>99 \%$ for $\delta / c \gtrsim 10^{2.8}$ years, given that $\mathcal{N} = 1$. Lastly, we list this toy model's significant limitations, and suggest ways to ameliorate them in more realistic future models.
\end{abstract}

\maketitle



\section{Introduction}
The Fermi paradox is the observation that there is an apparent contradiction between the lack of obvious evidence for the existence of extraterrestrial life, and the seemingly likely multitude of habitable planets in our own galaxy. This can be quantified via the Drake equation, which can imply implausibly large numbers of communicative civilizations even with arguably plausible values of the input parameters. This apparent contradiction makes the \citep{morrison1979search} Search for Extraterrestrial Intelligence (SETI) all the more interesting, since the passive search for life-produced radio signals has thus far not clearly yielded any such signals. This has, at least in part, motivated a debate regarding whether or not SETI is a useful exercise (see \cite{bradbury2011dysonian,cir2013seti}). This paper will not directly attempt to answer this question, but will, however, suggest a general framework within which this question can be quantified. \par
As motivation, we can begin with a general question. If there are currently $N$ signals from communicative civilizations in our galaxy, how likely is it that one would have been observed since the start of SETI? Suppose that we consider values of $N$ that seem reasonably conservative, but we find that the probability of having found a signal to date from at least one of these sources is vanishingly small, given some simplifying assumptions; one may then argue that passive searches may not be meaningful. If this probability is large, however, one may argue that a lack of results from SETI is scientifically significant. In this work, we will explain the general conditions under which the latter scenario prevails. Further, we will explain how one can use such a result to constrict the parameter space of the Drake equation. This is in contrast to the study of the Drake equation in \citep{prantzos2020probabilistic1}, in which the parameter space of the Drake equation is constrained via civilizations physically expanding their spheres of influence throughout the galaxy. \par
The reader may wonder what we mean by ``some simplifying assumptions''. We define and discuss several such assumptions, and we will leave it to the reader to decide their plausibility. We hope that this paper will provide a framework in which to consider interpretations and resolutions of the Fermi paradox, within the context of past, present, and future searches. Here, we will refer to ``communicative civilizations'' as ``civilizations''. \par
In the next section, we discuss the conceptual crux of this paper. We then conceptually introduce our  geometrical model of the MW galaxy. This is followed by a discussion of how to compute the probability of at least one observation from the probability of individual observations, and a discussion of how one can use these results to constrain the Drake equation. Lastly, we interpret our work within the context of previous SETI and Fermi paradox research, and provide suggestions for fixing the limitations of the simple model described here. \\

\subsection{Constraining the Drake Equation via SETI}

The essential idea we wish to present is that a null SETI result can be used to constrain the Drake equation (\cite{drake1965radio})

\begin{equation}\label{Drake}
N = r_* f_p n_e f_l f_i f_c l \equiv \mathcal{N}l,
\end{equation}

\noindent where 

\begin{itemize}
\item $r_*$ is the number of stars forming per year in the Milky Way galaxy;
\item $f_p$ is the fraction of such stars with planets;
\item $n_e$ is the number of habitable planets, per planetary system;
\item $f_l$ is the fraction of habitable planets on which life evolves;
\item $f_i$ is the fraction of life-bearing planets on which intelligent life evolves;
\item $f_c$ is the fraction of intelligent lifeforms which emit EM radiation;
\item $l$ is the mean number of years during which intelligent lifeforms emit EM radiation.
\end{itemize}

\noindent The likelihood $P_i$ of observing a civilization's signals depends on the number of years during which the civilization has emitted signals. We can then determine the probability $\mathcal{P}$ of observing \textit{at least} one signal by imagining many civilizations. Therefore, the fact that SETI has not observed any life-produced signals can be used to place constraints on $l$. By combining an upper bound on $l$ with Equation \ref{Drake}, we can exclude parts of $\mathcal{N}-l$ parameter space, and, in this way, a null SETI result quantifies the Fermi paradox. \par
So as not to misguide the reader by focusing on mathematical details, we will give a brief and conceptual introduction to our geometrical model in the next subsection, sufficient to follow the connection between $P_i$ and the Drake equation. The details are provided in Appendix \ref{appedix}.

\subsection{Conceptual introduction to the geometrical model}\label{ConceptualIntroduction}

\begin{figure}[h]
	\centering
	\includegraphics[width=1.00\linewidth]{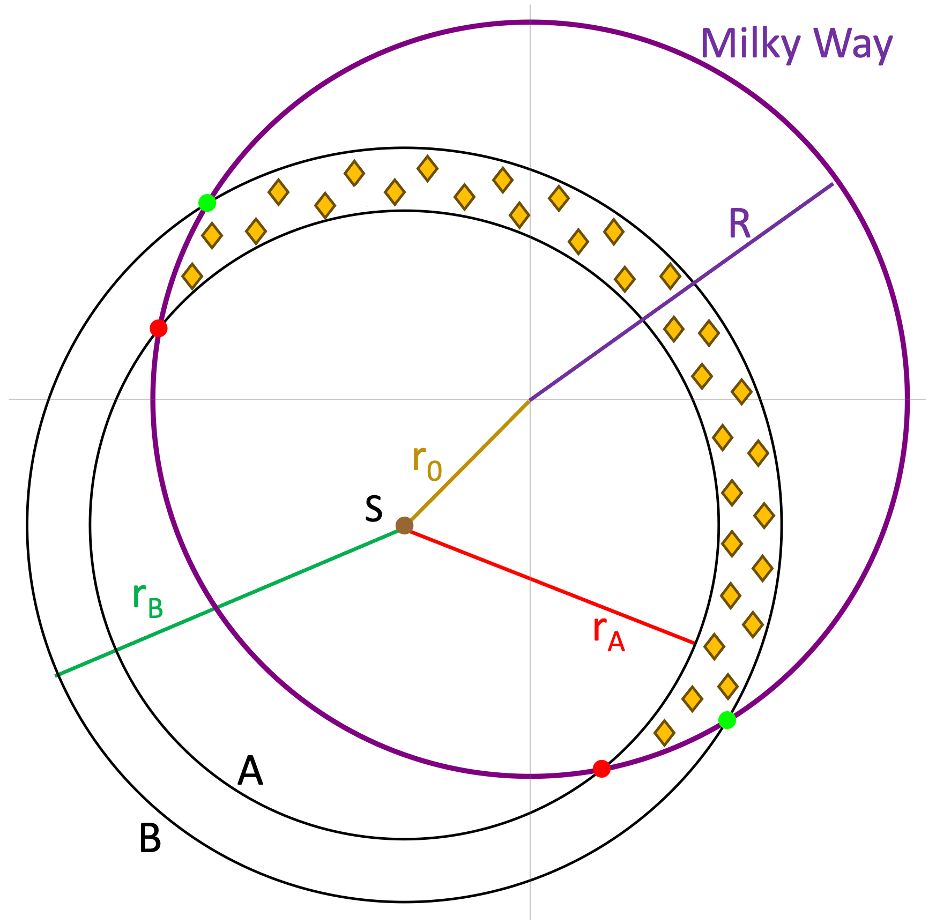}
	\caption{\footnotesize This figure conceptually describes the geometrical model used in this paper. The essential idea is that the greater the number of years during which a civilization has emitted signals, the greater the likelihood of observing those signals. This likelihood is proportional to the signal area, which is denoted with yellow diamonds. Here, the signal is being sent from a source ``$S$'', which is a distance ``$r_0$'' from the center of the MW. The first and last signals have radii $r_B$ and $r_A$. Thus, $l$ in Equation \ref{Drake} can be connected to the probability $\mathcal{P}$ of observing at least one signal.}\label{fig:Conceptual}
\end{figure}

\indent In Figure \ref{fig:Conceptual}, the basic features of our geometrical model are depicted. We model the MW as a two-dimensional disk of radius $R$, containing civilizations (labeled ``$S$'') located at distances $r_0$ from the center of the MW. The first and last signals have radii $r_B$ and $r_A$, respectively. The probability of observation of a single signal ($P_i$) is proportional to the signal area, denoted by yellow diamonds in the aforementioned figure. This area is derived in Appendix \ref{appedix}. \par
\indent For some applications, the exact area is not needed; in such applications, $r_B - r_A \equiv R \delta \ll R$\footnote{From here on, we will normalize $\delta$ as $\left(r_B - r_A \right) / R$ and take $R = 1$, redefining $R$ in units of distance only as necessary.}. Since our primary goal in this paper is to conceptually describe how SETI can be used to constrain the Drake equation, we will for simplicity take $\delta \ll 1$ throughout. In Appendix \ref{appedix}, we show that in this regime the probability of observing a signal is $P_i = 0.6 \delta$. \par
\indent In Table \ref{probtable}, we list the primary parameters used in the paper, summarize their meanings, and direct the reader to the equation or section in which they are defined. Note that there are three ways in which we will compute $P_i$. The aforementioned approximation for $P_i$ when $\delta \ll 1$ we call $P_{\delta}$. Contrarily, the exact probability (Equation \ref{Pdef}) we refer to as $P_{exact}$. Finally, $P_{test}$ is calculated by iteratively generating sources, and explicitly determining if each source is observable. This is discussed in Section \ref{numerical}. Next, we will discuss how to connect $N$, $P_i$, and $\mathcal{P}$.

\begin{table}[]
\caption{\footnotesize Here we summarize all of the primary symbols we use in this paper, and provide links to the sections in which they are defined.}\label{probtable}
\centering
\begin{tabularx}{\linewidth}{| >{\centering\arraybackslash}m{1.2cm} || L ||  >{\centering\arraybackslash}m{1.2cm} ||}
\hline
Symbol  & Meaning   & See  \\
\hline
$\vspace{0.12cm} l$       & mean lifetime of civilizations (years), $= r_B -r_A$, where $c = 1$, for a single civilization   & Section \ref{FermiSETI} \vspace{0.04cm}  \\
\hline
$\vspace{0.04cm} \delta$       & $\vspace{0.04cm} l/R$   & Section \ref{ConceptualIntroduction}   \\
\hline
$\vspace{0.08cm} P_{exact}$       & the exact probability of a single observation   & Equation \ref{Pdef}   \\
\hline
$\vspace{0.05cm} P_{\delta}$       & $\vspace{0.04cm} 0.6\delta$   & Equation $\ref{AdeltaAve}$   \\
\hline
$\vspace{0.3cm} P_{test}$       & the probability of a single observation, via iterating through all EM sources and directly checking the observation condition  & Equation \ref{Ptestdef} \vspace{0.19cm} \\
\hline
$\vspace{0.05cm} \mathcal{P}$       & the probability of at least one SETI observation  & Section \ref{IntroProb}    \\
\hline
$\vspace{0.05cm} \mathcal{N}$       & all factors in the Drake equation except $l$  &  Sections \ref{FermiSETI}   and \ref{Drake} \\
\hline
\end{tabularx}
\end{table}

\section{Connecting SETI to our Geometrical Model}
In this section, we will discuss how to connect a null SETI result to $P_i$ and $\mathcal{P}$. We will then simplify our discussion by taking $\delta \ll 1$, and by considering only a homogeneous distribution of $P_i$.

\subsection{Probability distributions}\label{IntroProb}

In this subsection, we discuss how to determine the probability of at least one detection $\mathcal{P}$ as a function of the distribution $P_i$. This is similar to the discussion in \citep{grimaldi2017signal}. For $N$ civilizations having independent detection probabilities $P_1$, $P_2$, ...,$P_N$, the probability of no detection is 

\begin{equation*}
P_{none} = \left(1 - P_1 \right) \cdot \left(1 - P_2 \right)...\left(1 - P_N \right) = \prod_{i=1}^N \left(1 - P_i \right),
\end{equation*}

\noindent and, thus, the probability of at least one detection is

\begin{equation}\label{ProbOneDetGeneral}
\mathcal{P} = 1 - \prod_{i=1}^N \left(1 - P_i \right). 
\end{equation}

\noindent Since we will eventually wish to explore how the model depends on $N$, we may write the above expression as 

\begin{equation}\label{ProbOneDetSum}
	\ln \left( 1- \mathcal{P} \right) = \ln \left( \prod_{i=1}^N \left(1 - P_i \right) \right) = \sum_i^N \ln \left(1 - P_i \right).
\end{equation}

\noindent One may analyze the probability distributions $P_i$ as in \cite{kreifeldt1971formulation} and \cite{wallenhorst1981drake}. Since our goal is to provide a highly general introduction to our model, however, we take the greatly simplified case where each probability $P_i = P$ is identical. We refer to this as the homogeneous probability model, which we can interpret as a toy model.

\subsection{The homogeneous probability model}

In this section, we will assume that each signal has an equal probability of observation. The probability of at least one detection is therefore

\begin{equation*}
\mathcal{P} = 1 - \left(1 - P \right)^N.
\end{equation*}

\noindent Solving for $\log_{10} N$, we obtain 

\begin{equation}\label{PNrel}
	\log_{10} N = \log_{10} \left[ \frac{\log_{10} \left( 1 - \mathcal{P} \right)}{\log_{10} \left( 1 - 10^\sigma \right)} \right],
\end{equation}

\noindent where $\sigma = \log_{10} P$. We now discuss how to approximate this relation in the $\delta \ll 1$ regime. 

\subsection{Approximating $\mathcal{P}$ in the $\delta \ll 1$ regime.}

\begin{figure}
	\centering
	\includegraphics[width=1.00\linewidth]{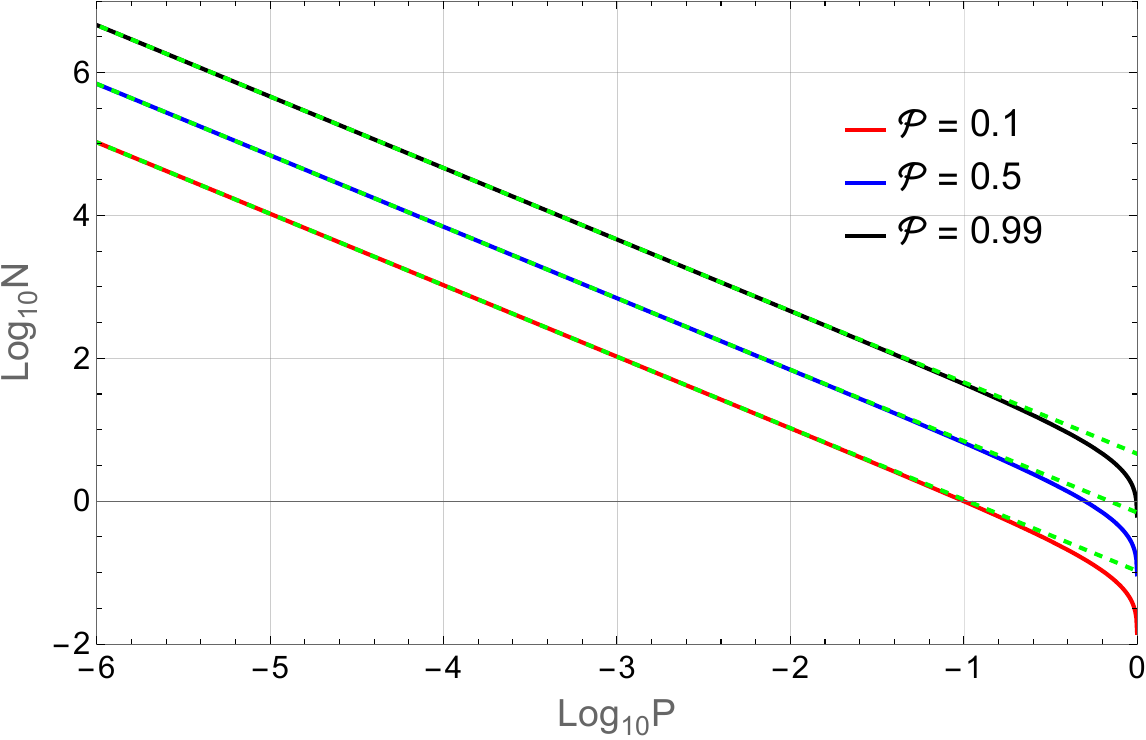}
	\caption{\footnotesize Here, we depict $\log_{10} N$ vs. $\log_{10} P$ (Equation \ref{PNrel}) in the homogeneous probability model, for $\mathcal{P} = 0.1$ (in red), $0.5$ (in blue), and $0.99$ (in black). These curves are overlayed with the small-$P$ approximation Equation \ref{PNrelSmallP}, shown in green dashed lines, for each $\mathcal{P}$.}\label{fig:ProbNFigure}
\end{figure}

In the $\delta \ll 1$ regime, the probability of a single observation $P$ will be $\ll 1$. This conclusion follows from Section \ref{smalldeltaregime}. We can now approximate Equation \ref{PNrel} in this regime, starting from Equation \ref{ProbOneDetSum}. This yields \par

\begin{align}\label{PNrelSmallP}
&\ln \left(1 - \mathcal{P} \right) = \sum_i^N \ln \left(1 - P_i \right) = N\ln \left( 1 - P \right) \approx -NP, \nonumber \\
&\Rightarrow \log_{10} N \approx - \log_{10} P +\log_{10} \left( \ln \left( 1 - \mathcal{P} \right)^{-1} \right),
\end{align}

\noindent where in the first line we have assumed that $P \ll 1$. This result is superposed on the exact result (Equation \ref{PNrel}) in Figure \ref{fig:ProbNFigure}, in which it is depicted as green dashed lines for $\mathcal{P} = 0.1$, $0.5$, and $0.99$. Equation \ref{PNrel} is plotted via the solid curves in red, blue, and black for the same $\mathcal{P}$ values. From this figure, we can see that Equation \ref{PNrelSmallP} is accurate to at least an order of magnitude so long as $\delta \lesssim 1$. \par
In the next section, we will discuss our numerical results; then, we discuss the implications for the Drake equation and Fermi paradox.

\vspace{0.75cm}

\section{Numerical Analysis of Our Geometrical Model}\label{numerical}

In Table \ref{probtable}, we publish the probability of a single observation, computed using three different methodologies\footnote{The Fortran code used to compute $r_{test}$ can be found \href{https://sites.google.com/view/mciviletti/research/r_test-code}{here}.}: $P_{exact}$, via Equation \ref{Pdef}; $P_{\delta}$, via Equation \ref{AdeltaAve}; and, $P_{test}$, via Equation \ref{Ptestdef}. The first is computed by splitting the domains of $r_0$ and $r_A$ into $10^3$ segments each for $\delta = 10^{-1},10^{-1}$, and $10^{-3}$, and $10^4$ segments each for $\delta = 10^{-4}$. The second is trivially derived. The third is calculated by splitting the domains of $r_0$, $r_A$, and $\theta_0$ into $10^3$ segments each for $\delta = 10^{-1},10^{-1}$, and $10^{-3}$, and $10^4$ segments for $\delta = 10^{-4}$. This also requires one to choose $r_E$ and $\theta_E$ values, which are the radius and angle of the observer (i.e., ``Earth)''\footnote{$\theta_0$, $r_E$, and $\theta_E$ are defined in Figure \ref{fig:CommunicationGeometry}.}; these, however, do not affect $P_{test}$. See Section \ref{smalldeltaregime} for a discussion. At each iteration, we check explicitly that the observation condition (Equation \ref{Ptestdef}) is met. The percent error in Table \ref{probtable} refers to the error of $P_{\delta}$ with respect to $P_{exact}$.

\begin{table}[]
\caption{\footnotesize Below we list observation probabilities computed three different ways. $P_{test}$ is derived from Equation \ref{Ptestdef}, computed by iterating $r_0$, $r_A$, and $\theta_0$ values over their domains and checking to see if the observation condition is met. See Figure \ref{fig:CommunicationGeometry} for the definitions of these parameters. P$_{exact}$ is computed by inserting iterated $r_0$ and $r_A$ values into Equation \ref{Pdef}, while P$_{\delta}$ is computed trivially via Equation \ref{AdeltaAve}. The percent error in Table \ref{probtable} refers to the error of $P_{\delta}$ with respect to $P_{exact}$.}
\begin{center}
\begin{tabular}{|c || c| | c || c || c ||}
\hline
$ \delta$  & $P_{exact}$   & $P_{\delta}$  & $\%$ error  & $P_{test}$ \\
\hline
$10^{-1}$       & $0.632526$   & $0.600000$   & $5.1$  & $0.636311$    \\
\hline
$10^{-2}$       & $0.603640$   & $0.600000$   & $5.8 \times 10^{-1}$  & $0.603781$   \\
\hline
$10^{-3}$       & $0.600549$  & $0.600000$   & $8.7 \times 10^{-2}$  & $0.600724$     \\
\hline
$10^{-4}$       & $0.600055$  & $0.600000$   & $3.4 \times 10^{-2}$   & $0.599790$   \\
\hline
\end{tabular}\label{probtable}
\end{center}
\end{table}

We now turn to the meaning of our model in the context of the Fermi paradox and SETI. \par

\section{Discussion}

In this section, we will apply our geometrical model to the Drake equation and discuss its implications. Finally, we will discuss the limitations of the toy model we present here.

\subsection{The Fermi paradox in light of SETI}\label{FermiSETI}

Since $P \approx 0.6 \delta$ for small $\delta$, we can write Equation \ref{PNrelSmallP} as

\begin{equation*}
\log_{10} N \approx - \log_{10} 0.6 \delta + \log_{10} \left( \ln \left( 1 - \mathcal{P} \right)^{-1} \right).
\end{equation*}

\noindent We can now write $\delta = lc/R$, where $c$ is the speed of light and $R$ is the radius of the MW galaxy in light-years. We will use years and light-year interchangeably, thus taking $c = 1$, and we take $R = 5 \times 10^{4}$. Thus\footnote{Recall that in Section \ref{ConceptualIntroduction} we took $c=1$ and $R \delta = r_B - r_A$; therefore, $\delta$ is the normalized distance that the EM signals have travelled through the MW galaxy, consistent with our result here.}, $\delta = \frac{1}{5} \cdot 10^{-4} l$. This gives

\begin{align*}
\log_{10}N &\approx  -\log_{10} \left(\frac{3}{25} \cdot 10^{-4} \right) - \log_{10} l + \log_{10} \left( \ln \left( 1 - \mathcal{P} \right)^{-1} \right), \\
&\approx  -\log_{10} l + \log_{10} \left(\frac{25}{3} \cdot 10^{4} \cdot \ln \left( 1 - \mathcal{P} \right)^{-1} \right) \\
&\equiv  -\log_{10} l + \mathcal{C} (\mathcal{P}).
\end{align*}

\noindent We may now combine this with Equation \ref{Drake}; this yields 

\begin{align}\label{curlyN-l}
\log_{10}N &= \log_{10} \mathcal{N} + \log_{10} l \approx -\log_{10} l + \mathcal{C} (\mathcal{P}), \nonumber \\
& \Rightarrow \log_{10}\mathcal{N} \approx \mathcal{C} (\mathcal{P}) - 2 \log_{10} l.
\end{align}

\begin{figure}
  \centering
  \includegraphics[width=1.05\linewidth]{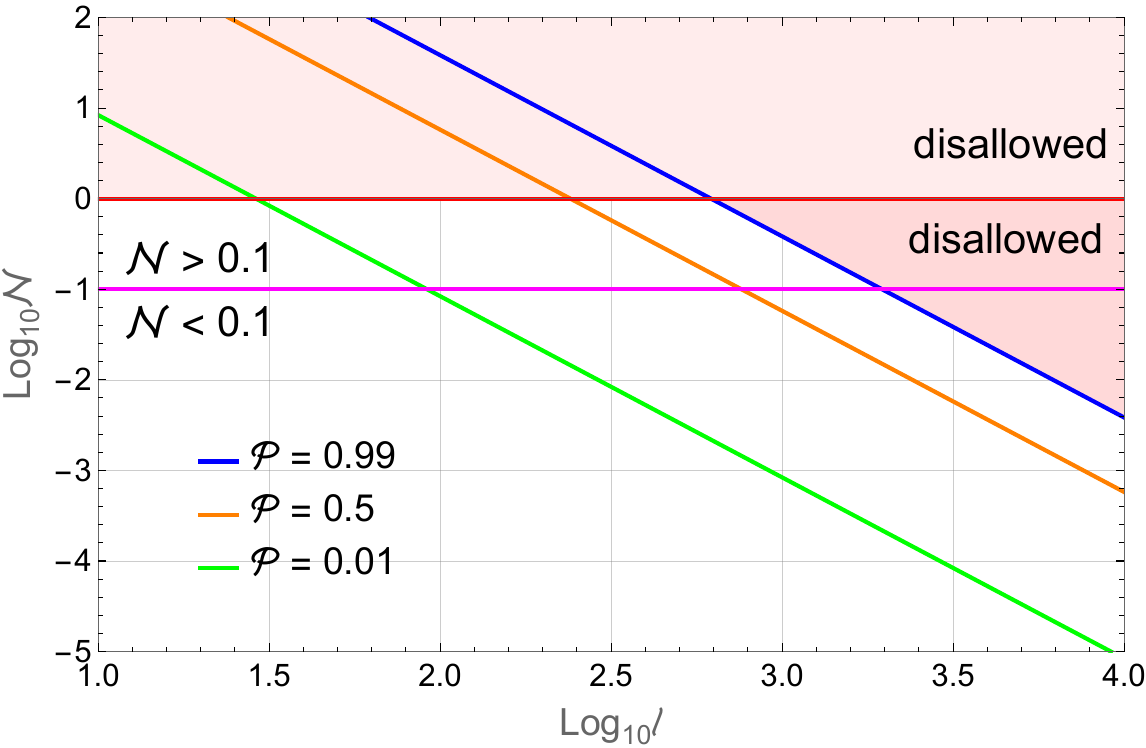}
\caption{\footnotesize Here, we plot $\log_{10} \mathcal{N}$ vs. $\log_{10} l$ (Equation \ref{curlyN-l}) for $\mathcal{P} = 0.99$, $0.5$, and $0.01$ (blue, orange, and green, respectively). These curves assume that $\delta \ll 1$. We also plot $\mathcal{N} \approx 0.1$ and $\mathcal{N} \approx 1$ (magenta and red horizontal lines, respectively). The rightmost part of the figure corresponds to $\log_{10} l = 4$, implying that $\log_{10} \left( \frac{1}{2} \cdot 10^5 \delta \right) = 4 \Rightarrow \delta < 1/5$ in this figure.}\label{fig:CurlyN-lplot}
\end{figure}

\noindent Let us now presume that $\mathcal{N} \lesssim 1$, as argued in the original formulation of the Drake equation \cite{drake1965radio}. This precludes the area above the red line in Figure \ref{fig:CurlyN-lplot}, in which we plot Equation \ref{curlyN-l}. The magenta line represents $\mathcal{N} = 0.1$, added for readability. The area to the top right of the blue line in that figure implies $\mathcal{P} > 0.99$, since the blue line represents $\log_{10}\mathcal{N}$ vs $\log_{10}l$ for $\mathcal{P} = 0.99$. Suppose we restrict ourselves to SETI results which have $>1 \%$ probabilities of occurring; this precludes the aforementioned area. In the original iteration, Drake considered $l = 10^3$ to $10^4$ years as reasonably conservative values. We can see from Figure \ref{fig:CurlyN-lplot}, however, that, if each civilization emits EM signals for $l \approx 10^{2.8}$ years, there is about a $99 \%$ probability that we would have observed at least one of these signals, given $\mathcal{N} \approx 1$. The latter presumption reduces Equation \ref{curlyN-l} to 

\begin{equation}\label{curlyNequals1Curve}
\log_{10}l \approx \frac{1}{2} \mathcal{C}(\mathcal{P}).
\end{equation}

\begin{figure}
  \centering
  \includegraphics[width=1.0\linewidth]{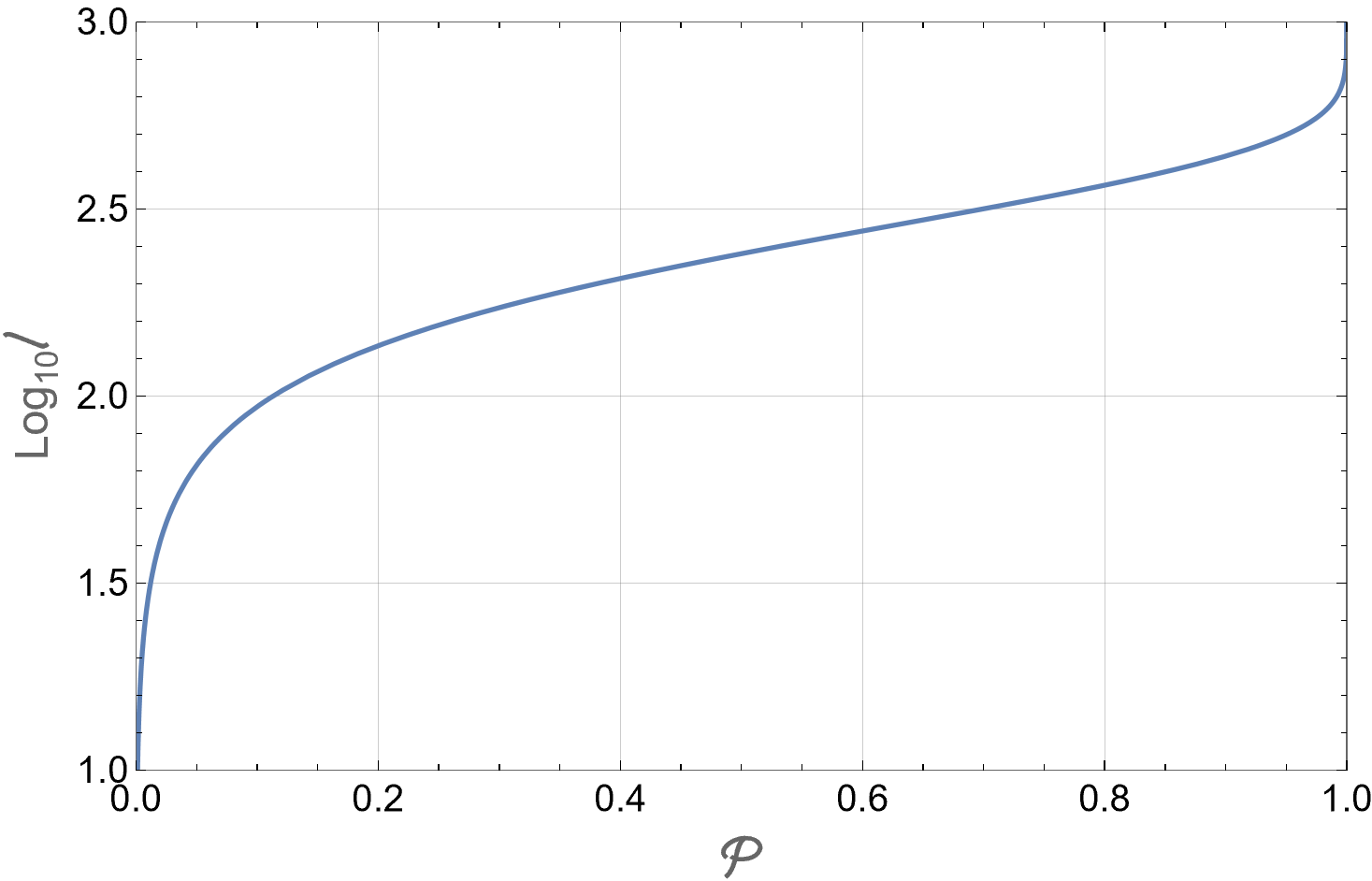}
\caption{\footnotesize Here, we plot $\log_{10} l$ vs. $\mathcal{P}$, as per Equation \ref{curlyNequals1Curve}. Thus, here we presume that $\mathcal{N} \approx 1$. This shows that, given the limitations of our toy model, it is unlikely that there are about $10^{2.8}$ civilizations or more unless $\mathcal{N}$ is considerably smaller than $1$. }\label{fig:lvscurlyP}
\end{figure}

These results, under the limitations described below, imply that it is unlikely that there are more than $\mathcal{O}(10^3)$ communicative civilizations unless $\mathcal{N}$ is appreciably smaller than $1$. This is depicted in Figure \ref{fig:lvscurlyP}. Since $\mathcal{N}$ is multivariate, this constraint cannot in and of itself restrict any of the individual parameters which define $\mathcal{N}$. There are, however, some independent constraints on some of those aforementioned parameters. For instance, $r_*$ is likely (\cite{elia2022star}) around $1$, and it currently seems likely that $f_p$ is (\cite{cassan2012one}) around $1/2$. If these hold, then $n_e f_l f_i f_c$ should likely be at least around an order of magnitude less than $1$ if $l  \gtrsim \mathcal{O}(10^3)$ and $\mathcal{N} \approx 1$. The model we have discussed here has rather notable limitations, and so this argument is far from dispositive. Perhaps with significantly more data and a more realistic model of this general type, however, more definitive statements along these lines might be made. Let us turn toward the meaning of what we've done in relation to similar research. 

\subsection{Comparing our work to previous SETI and Fermi paradox research}

Thus far we have not placed our work within the context of previous SETI and Fermi paradox research. What makes this paper unique is that we present a method by which the null SETI result can be used to impose constraints upon $l$; and, hence, we can constrain the parameter space of the Drake equation. Previous research (\cite{forgan2009numerical, maccone2012statistical, miller2017fermi}, \cite{maccone2012statistical}) has presented numerical methods for simulating the formation of life in the MW galaxy, via Monte Carlo techniques and Bayesian analysis. Similar work has been done within particular Fermi paradox frameworks--for instance, within the context of the Great Filter hypothesis (\cite{aldous2012great}). Statistical analysis has been used to argue that the Fermi paradox is not likely to be paradoxical at all (\cite {sandberg2018dissolving}). Others (\cite{westby2020astrobiological}) have investigated the impact of astrophysical scenarios on the Drake equation, constraining its parameters via astronomical observations. \par
In this work, we aim to fill a rather conspicuous gap in the literature, by showing that SETI can be a useful tool by which to analyze the prevalence of civilizations within our galaxy. Merely by searching for life-produced signals and failing to find any, we can provide a quantitative interpretation of the Drake equation that, when combined with independent evidence, may one day provide humanity with a scientific understanding of its place within the Galaxy. The model we have presented here is intended, however, to give an example of how one might accomplish this; we are thus not naive about its limitations. In this next section, we turn to the limitations of this model and suggestions for how to improve it.

\subsection{Limitations of this model}\label{LIM}

The reader may have noticed significant limitations of the model we've presented in this paper. Let us list them:

\begin{enumerate}
\item Our model is two-dimensional. We justified this because the ratio of the diameter of the MW galaxy to the thickness of its disk is about $100$. Nonetheless, a three-dimensional model is more realistic.
\item We have not accounted for the attenuation of EM signals or the limitations of current radio wave observatories. Instead, we have assumed that signals of any magnitude may be observed, and therefore have not imposed any constraints on the locations of these signals. The power output of civilizations, as discussed in \cite{kardashev1964transmission}, is pertinent to providing a more realistic treatment. Future work should consider inhomogeneous distributions of civilizations by, amongst other variables, distance to Earth and power output.
\item We have not provided a mechanism to account for the density of EM sources. It may be the case, for instance, that civilizations may be more likely to be found in certain regions (i.e., the galactic habitable zone--see \cite{gowanlock2011model}) of the galaxy than others; our model presupposes that civilizations are equally likely to originate from any location in the galaxy. The general probability that at least one signal will be observed (Equation \ref{ProbOneDetGeneral}), however, in principle involves a distribution of probabilities $P_i$. In this paper we have assumed a homogeneous distribution $P_i = P$ of independent probabilities, so as to produce a toy model as an example of how our analysis may be applied.
\item Our analysis considers the existence of currently-observable EM signals within the galaxy, and assumes that all signal radii are equally likely. It may be the case, however, that the number of civilizations is time-dependent. This paper only considers the observation of signals last produced within about the past $10^5$ years, which is about five orders of magnitude less than the time during which the MW galaxy has existed. The distribution of observational windows, as well as stellar lifetimes and development times, is discussed in \cite{kreifeldt1971formulation,wallenhorst1981drake}. Future research may wish to apply these concepts to our model. 
\item We have assumed that a civilization emits signals continuously for a period of time $l/c$. Future research may want to consider other scenarios in which there could be intermittent emission.
\item In this paper, we have relied upon the approximation $\delta \ll 1$, and we have taken $P = 0.6 \delta$. A more realistic treatment might re-derive Equation \ref{curlyN-l}, for example, without assuming $P \approx \delta \ll 1$. Such a model would therefore consider signals emitted for of order 10,000 years or more.
\end{enumerate} 

\section{Conclusion}

We have articulated a model which can provide a quantitative assessment of the Fermi paradox. This model has serious limitations, but we argue here that it can provide a generic framework in which to quantify the usefulness of SETI in respect to the existence of intelligent life. Our model shows that the lack of observation of life-produced EM signals places limitations on the number of EM-emitting civilizations. Given the presumptions in our model, we show that there is about a $99 \%$ probability of observing at least one EM signal out of about $10^{2.8}$, if $\mathcal{N} = r_* f_p n_e f_l f_i f_c \approx 1$. Although this result is derived from a toy model, we have nonetheless shown that, at least in principle, one can rule out parts of $\mathcal{N}-l$ space merely from a null SETI result. As future research reduces the limitations discussed in Section \ref{LIM}, constraints on $\mathcal{N}-l$ space from the model presented in this paper may help provide insight into the resolution of the Fermi paradox.

\bigskip

\appendix

\section{A Simple Two-Dimensional Geometrical Model}\label{appedix}

Here, we derive the previously-cited geometrical results associated with our geometrical model. This discussion will be generally similar to ~\citet{grimaldi2017signal}, except that we will provide an exact solution in the two-dimensional case, and we will show that in the simplest interpretation the location of Earth does not affect the likelihood of observation. The latter provides a very useful way of generalizing these results to more complex probability distributions. Our treatment differs further in that we then connect these results to the Drake equation, and argue that its parameter space can be constrained via a null result from SETI.

\subsection{Introduction}\label{intro}

\begin{figure}
  \centering
  \includegraphics[width=1.0\linewidth]{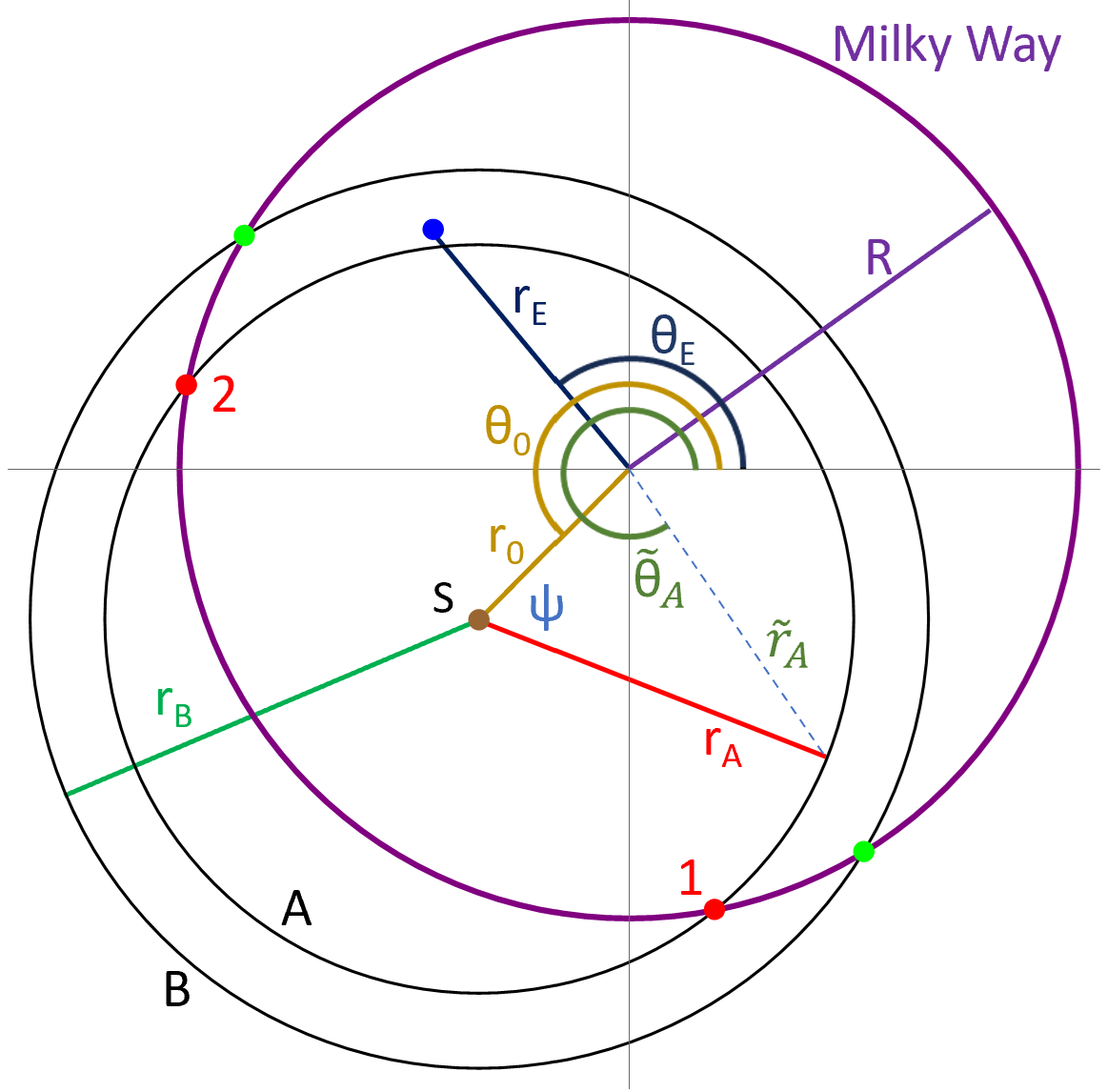}
\caption{\footnotesize Here, we depict the basic geometry of our model. The MW galaxy has a radius ``R''. The two light-fronts, emanating from an extraterrestrial source ``S'', are labelled ``A'' and ``B''. The source is a distance $r_0$ from the center of the MW galaxy, and $\tilde{r}_A$ is the distance from the origin to a point on light-front A. The distance of the Earth from the center of the Milky Way is labeled ``$r_E$'', and this figure depicts the condition defined by Equation \ref{Ptestdef}.}\label{fig:CommunicationGeometry}
\end{figure}

Since the ratio of the diameter of the Milky Way galaxy to the thickness of its disk is about $100$ (see \cite{goodwin1998relative} and \cite{de2010mapping}), we will approximate the galaxy as a two-dimensional disk. In Figure \ref{fig:CommunicationGeometry}, we depict a simplified disk model of the galaxy with two example light fronts emitted by a civilization; the former is represented by the purple circle, of radius $R$, and the latter two are represented by the two concentric black circles that are centered on the civilization ``S''. The radii of the light-fronts are $r_A = ct_A$ and $r_B= ct_B$, where $t_A$ and $t_B$ denote the amount of time that has passed since the emission of light signals $A$ and $B$. We assume that these correspond to the first and last signals, respectively, that the civilization sends. Further, we assume continuous transmission from $t_A$ to $t_B$. These and other assumptions contained here are discussed in Section \ref{LIM}. \par

These light fronts create an\footnote{The annulus is non-concentric; for simplicity, we will simply refer to it as an ``annulus''.} annulus whose area may only be partly located within the MW galaxy. Let us discuss how to compute this. The location of the source is $(r_0, \theta_0)$, and Earth is denoted by the blue dot, of position $(r_E, \theta_E)$. Observation will correspond to an orientation where the Earth is in-between the two light-fronts. Hence, observation will only occur when 

\begin{equation}\label{Ptestdef}
r_A < \sqrt{r_E^2 + r_0^2 - 2r_Er_0 \cos \left( \theta_E - \theta_0 \right)} < r_B. 
\end{equation}

For a single pair of light fronts, the probability of observation for a randomly-placed observer is the probability that this randomly-placed observer will fall in-between the light fronts; this is $\tilde{A}/ \pi R^2$, where $\tilde{A}$ is the area that is both in-between the two light fronts and also within the MW galaxy. The area $\tilde{A}$ is circumscribed by the curves whose intersections with the edge of the MW galaxy are defined by the red and green dots. All points along the light-front $A$ ($B$) form angles $\tilde{\theta}_A$ ($\tilde{\theta}_B$), as shown in Figure \ref{fig:CommunicationGeometry}. The distance from the center of the MW galaxy to any point along light-front $A$ ($B$) is $\tilde{r}_A$ ($\tilde{r}_B$). The red (green) dots are defined by $\tilde{r}_A = R$ ($\tilde{r}_B = R$), which implies that (keeping to just the red dots as an example)

\begin{equation}\label{rAdef}
r_A^2 = R^2 + r_0^2 - 2Rr_0 \cos (\theta_0 - \tilde{\theta}_A),
\end{equation}

\noindent and thus

\begin{equation}\label{thetaAdef}
\tilde{\theta}_A = \theta_0 \pm \arccos \left[ \frac{R^2 +r_0^2 - r_A^2}{2r_0R} \right] \equiv \theta_0 \pm \Delta \theta.
\end{equation}

\indent We can split useful $(r_0,r_A)$ parameter space into two parts: (1) where the argument of $\arccos$ is within $[-1,1]$, and (2) where it is not. In the first case we will have two\footnote{Except the singular case where  $\Delta \theta = 0$ when the argument is $1$.} values for $\tilde{\theta}_A$, which we will call $\tilde{\theta}_{A1} = \theta_0 - \Delta \theta$ and $\tilde{\theta}_{A2} = \theta_0 + \Delta \theta$. This implies that

\begin{equation}\label{rArestr}
R + r_0 > r_A > R - r_0.
\end{equation}

\noindent The upper limit is required so that at least part of the source circle is inside the MW, and the lower limit ensures that the source circle is not completely within the MW. We will refer to circles that satisfy both of these as partial source circles. If Equation \ref{rArestr} is not satisfied, then the second case applies and the circle will be completely inside the MW. We refer to these as whole source circles. Hence, in this scenario $r_A \leq R - r_0$ and $\tilde{A} = \pi (r_B^2 - r_A^2)$. In the next subsection, we will compute the area for both scenarios. 

\subsection{Annulus area and probability of observation}

\begin{figure}
	\centering
	\includegraphics[width=1.0\linewidth]{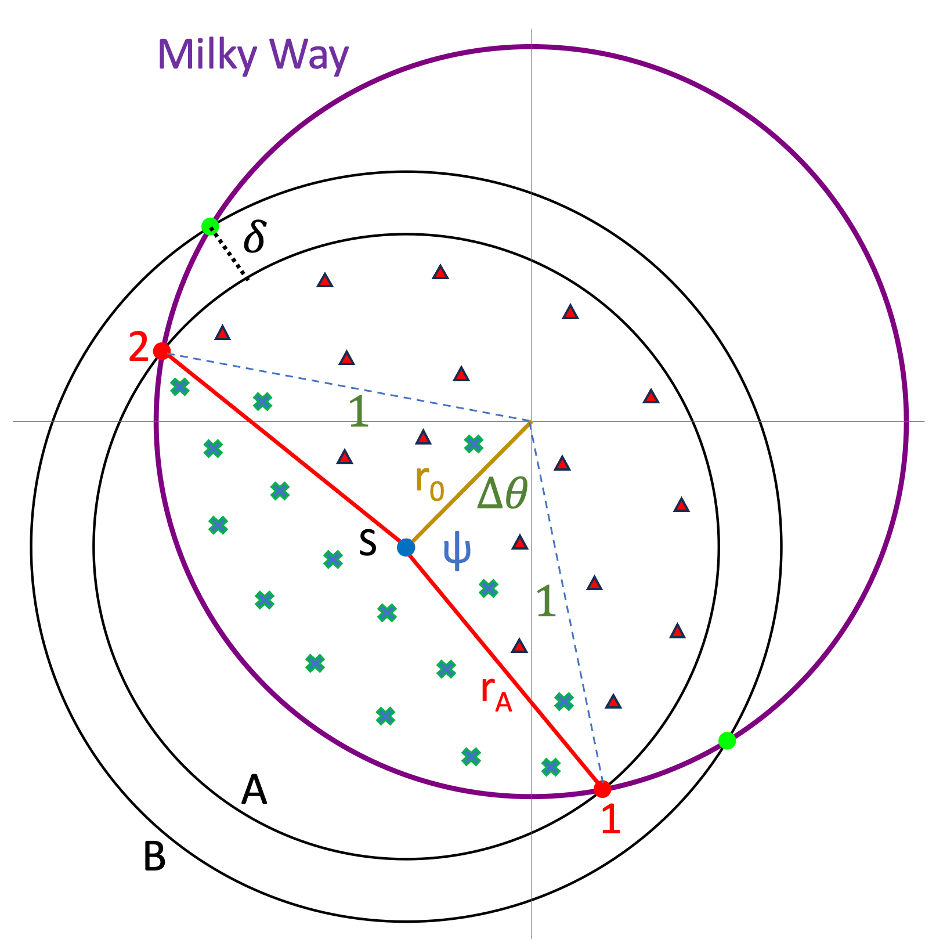}
	\caption{\footnotesize Here, we depict the three areas that compose $\tilde{A}$: the area denoted by red triangles ($\psi r_A^2$), the area denoted by green x symbols ($\Delta \theta$), and their overlap. Thus, we subtract the last area from the sum of the first two. Note that here we take $R=1$.}\label{fig:MWdiskModelIntermediateArea}
\end{figure}

In Figure \ref{fig:MWdiskModelIntermediateArea}, we depict just the essential components of our geometrical model sufficient to compute  $\tilde{A}$. The inner area, $\tilde{A}_A$, can be bifurcated into two parts: the area along the inner concentric circle from one red dot to the other (shaded with red triangles), and the area along the edge of the MW (shaded with green x marks) from one red dot to the other. The latter is $R^2 \left( \tilde{\theta}_{A2} - \tilde{\theta}_{A1} \right)/2 = R^2 \Delta \theta = \Delta \theta$, taking $R = 1$; the former is $r_A^2 \psi$, where $\psi$ is (as depicted in Figures \ref{fig:CommunicationGeometry} and \ref{fig:MWdiskModelIntermediateArea}) the angle between $r_0$ and $r_A$; $\psi$ appears to be 

\begin{align}\label{PsiDef}
	\nonumber
    \frac{\sin \psi}{\tilde{r}_A} &= \frac{\sin \Delta \theta}{r_A}, \\ \nonumber
    \Rightarrow \psi & = \sin^{-1} \left( \frac{\sqrt{4r_0^2 - \left( 1 + r_0^2 - r_A^2 \right)^2}}{2 r_0 r_A} \right) \equiv \sin^{-1} \left( \chi \right),
\end{align}

\noindent where we have taken $\tilde{r}_A = 1$ so that $\psi$ is defined as in Figure \ref{fig:MWdiskModelIntermediateArea}. This angle isn't quite right, however, since $\chi$ approaches $0$ as $r_A$ approaches\footnote{As $r_A$ approaches $1-r_0$, $1 + r_0^2 - r_A^2$ approaches $2 r_0$; therefore the
numerator in the definition of $\chi$ vanishes.} $1 - r_0$. As we can see from Figure \ref{fig:MWdiskModelIntermediateArea}, however, $\psi$ should be $\pi$ in this limit. Thus,  

\begin{equation}\label{PsiDef}
   \psi = \gamma \pi + \alpha \sin^{-1} \left( \chi \right),
\end{equation}

\noindent where $\alpha = -1$ ($\gamma = 1$) when $\psi > \pi/2$ and $+1$ ($0$) otherwise. When these two regions meet at $\psi = \pi/2$, 

\begin{equation*}
	\sin^{-1} \left( \chi \right) = \pi - \sin^{-1} \left( \chi \right),
\end{equation*}

\noindent which implies that $\chi = 1$ and 

\begin{equation*}
    r_A^2 + r_0^2 + 1 - 2r_A - 2r_0 + 2r_Ar_0 = 0.
\end{equation*}

\noindent Equation \ref{rAdef} implies that $r_A = \sqrt{1 + r_0^2 - 2r_0 \cos \Delta \theta}$, which, when input into the equation just above, gives $r_0 = \cos \Delta \theta$. Thus, $r_A = \sqrt{1 + r_0^2 - 2r_0^2} = \sqrt{1 - r_0^2}$ when\footnote{We can see this also via the Pythagorean theorem, since the triangle in Figure \ref{fig:MWdiskModelIntermediateArea} is a right triangle when $\psi = \pi/2$.} $\psi = \pi/2$. Thus, $\alpha = -1$ ($\gamma = 1$) when $r_A < \sqrt{1 - r_0^2}$ and $+1$ ($0$) otherwise. \par
As we've mentioned, the inscribed area can be split into two areas: $\Delta \theta$ and $r_A^2 \psi$. These two areas, however, overlap. We can see that, from Figure \ref{fig:MWdiskModelIntermediateArea}, this overlapping area (denoted by green x marks and red triangles) is $0.5 r_0 \sin \Delta \theta$. Thus, we must subtract $r_0 \sin \Delta \theta$ from $\Delta \theta + r_A^2 \psi$. The total area is thus

\begin{equation}\label{areaAtilde}
\tilde{A}_A = \Delta \theta + r_A^2 \psi - r_0 \sin \Delta \theta.
\end{equation}

\noindent Therefore, the probability of a single observation is

\begin{equation}\label{Pdef}
P =  \begin{cases}
	\left(\tilde{A}_B - \tilde{A}_A \right)/\pi, \hspace{0.5cm} 1 + r_0 > r_A > 1 - r_0, \\
	r_B^2 - r_A^2, \hspace{1.3cm} r_A \leq 1 - r_0, \\
\end{cases}	
\end{equation}

\noindent where $\tilde{A}_B$ is identical to $\tilde{A}_A$, except that the radius of the source circle is $r_B = r_A + \delta$.

\vspace{0.75cm}

Throughout this paper, we explore the mechanics of this approach via a toy model. This model is based upon several approximations, and, thus, the exact results are superfluous. Since for many practical purposes $r_B - r_A$ will be small compared to the radius of the MW, for example, we will in the next section implement this approximation and use it throughout the paper.

\subsection{$\delta \equiv r_B - r_A \ll 1$ regime} \label{smalldeltaregime}

In some realistic scenarios, the distances between initial and final EM signals may be at least an order of magnitude less than the radius of the MW. In this section, we will proceed under this assumption. Approximations in this regime will be labeled with a subscript $\delta$; for example, the left side of Equation \ref{Pdef} in this regime would be written $P_\delta$. \par
Let us now approximate Equation \ref{Pdef} where $\delta \ll 1$. We begin with the bifurcation of the parameter space into whole source circles and partial source circles, as described in section \ref{intro}. When $\delta \equiv r_B - r_A \ll 1$, we may approximate the area $\tilde{A}$ as $\tilde{A} \approx \Omega \cdot \delta r_A$, where $\Omega = 2 \pi$ when the annulus is completely within the MW and $\Omega = 2 \psi$ otherwise. See Equation \ref{PsiDef} for the definition of $\psi$. Thus,

\begin{align}\label{Adelta}
\nonumber
\tilde{A} \approx A_\delta &= r_A \delta \bigg( 2\psi \cdot \zeta [rA - \left( 1 - r0 \right)] \zeta [-rA + \left( 1 + r0 \right)] \\ 
&+ 2 \pi \cdot \zeta [-rA + 1 - r0] \bigg),
\end{align}

\noindent where

\begin{equation}\label{deltaphidef}
\zeta[x] =  \begin{cases}
1, \hspace{1cm} x > 0, \\
0, \hspace{1cm} x < 0, \\
\end{cases}
\end{equation}

\noindent is the Heaviside step function. The expression $\zeta [rA - \left( 1 - r0 \right)] \zeta [-rA + \left( 1 + r0 \right)] = \zeta_p$ enforces the first conditional in Equation \ref{Pdef} (i.e., partial circles), and $\zeta [-rA + 1 - r0] = \zeta_w$ enforces the second in Equation \ref{Pdef} (i.e., whole circles). The average area over the MW is, therefore,

\begin{align}\label{AdeltaAve}
\nonumber
\bar{A}_\delta &= \left( \int_0^2 \int_0^1 r_0 \left( \zeta_p + \zeta_w \right) dr_0 dr_A \right)^{-1} \\ \nonumber
&\times \int_0^2 \int_0^1 r_0 A_\delta dr_0 dr_A \\ \nonumber
&= 0.6 \pi \delta, \nonumber \\
\Rightarrow P \approx P_{\delta} &= 0.6 \delta, 
\end{align}

\noindent where in the last step we use the fact that the probability of observing a signal is the ratio of the area of the signal inscribed by the MW to the area of the MW. 

\subsection{The effect of the observer's location}
This result is averaged over all sources and observers. Since all SETI data corresponds to the location of the Earth within the MW, however, one might wonder about the relationship between the probability of an observation and the observer's location. This can be determined via Equation \ref{Ptestdef}, which shows that the probability is independent of $r_E$. If we solve that equation such that observation occurs only when $r_A \leq 1 + r_0$, we obtain

\begin{equation}
r_0 \geq \frac{-1 + r_E^2}{2 \left(1 + r_E \cos ( \theta_E - \theta_0 ) \right)},
\end{equation}

\noindent which is satisfied for all $r_E$. We can go further and use this to recompute Equation \ref{AdeltaAve}. The probability of observation is the likelihood that an observer will fall in-between the light fronts $r_A$ and $r_A + \delta$, and, since the domain of $r_A$ is $[0,1+ r_0]$, this probability is $\delta/\left( 1+ r_0 \right)$ if we assume that all $r_A$ are equally likely. Defining $\zeta = \zeta [-rA + \left( 1 + r_0 \right)]$, the average probability for small $\delta$ is therefore

\begin{align}\label{Pdirect}
\nonumber
P_\delta &= \left( \int_0^2 \int_0^1 r_0 \zeta dr_0 dr_A \right)^{-1} \times \int_0^2 \int_0^1 \frac{\delta r_0}{1 + r_0} \zeta dr_0 dr_A \\ \nonumber
& = 0.6 \delta.
\end{align}

\noindent This can be naturally generalized to inhomogeneous probability distributions, the integrand being $\delta \zeta r_0 P\left( r_0, r_A \right)/\left( 1+ r_0 \right)$.

\bibliographystyle{mnras}
\bibliography{refs} 


\end{document}